\documentclass[cup7a]{cupbook}

\usepackage{graphicx}

\title{Foundations of supernova cosmology}

\author[R.P. Kirshner]{Robert P. Kirshner}

\begin{document}

\pagenumbering{roman}
\maketitle
\tableofcontents
\cleardoublepage
\pagenumbering{arabic}

\chapter{Foundations of supernova cosmology}

\section{Supernovae and the discovery of the expanding universe}

Supernovae have been firmly woven into the fabric of cosmology from the very beginning of modern understanding of the expanding, and now accelerating universe.  Today's evidence for cosmic acceleration is just the perfection of a long quest that goes right back to the foundations of cosmology.  In the legendary Curtis-Shapley debate on the nature of the nebulae, the bright novae that had been observed in nebulae suggested to Shapley (1921) (see Trimble, 1995) that the systems containing them must be nearby.  Otherwise, he reasoned, they would have unheard-of luminosities, corresponding to M = -16 or more.  Curtis (1921) countered concluding that ``the dispersion of the novae in spirals and in our galaxy may reach ten magnitudes...a division into two classes is not impossible.''  Curtis missed the opportunity to name the supernovae, but he saw that they must exist if the galaxies are distant.  Once the distances to the nearby galaxies were firmly established by the observation of Cepheid variables (Hubble, 1925), the separation of ordinary novae and their extraordinary, and much more luminous super cousins, became clear.  

A physical explanation for the supernovae was attempted by Baade and Zwicky (1934). Their speculation that supernova energy comes from the collapse to a neutron star is often cited, and it is a prescient suggestion for the fate of massive stars, but not the correct explanation for the supernovae that Zwicky and Baade studied systematically in the 1930s.  In fact, the spectra of all the supernovae that they discovered and followed up in those early investigations were of the distinct, but spectroscopically mysterious, hydrogen-free type that today we call SN Ia.  They are not powered by core collapse, but by a thermonuclear flame. Baade (1938) showed that the luminosities of the supernovae in their program  were more uniform than those of galactic novae, with a dispersion of their peak luminosities near 1.1 mag, making them suitable as extragalactic distance indicators.  Right from the beginning, supernovae were thought of as tools for measuring the universe.

Nature has more than one way to explode a star.  This was revealed clearly by Minkowski (1941) who observed a distinct spectrum for some supernovae, different from those obtained for the objects studied by Baade.  SN 1940B had strong hydrogen lines in its spectrum.  These are the stars whose energy source we now attribute to core collapse in massive stars.  At the time, it seemed sensible to call Baade's original group Type I (SN I) and the new class Type II (SN II).   The small dispersion in luminosity for Baade's sample resulted from his good luck in having  Zwicky discover a string of supernovae that were all of a single type.  SN I are generally less luminous than the galaxies in which they occur. (Introductory texts, and introductory remarks in colloquia concerning supernovae usually get this basic fact wrong.) The SN II are, generally speaking, fainter than SN I and have a larger dispersion in their luminosity.  Separating the supernovae, on the basis of their spectra, into distinct physical classes is one way they have become more precise as distance indicators.  By the late sixties, Kowal (1968) was able to make a Hubble diagram for 19 SN I.  The scatter about the Hubble line for this sample, which reached out to the Coma Cluster of galaxies at a redshift of ~7000 km/s was about 0.6 magnitudes.   These were photographic magnitudes, obtained with the non-linear detectors of the time, and they contained no correction for absorption by dust in the host galaxies, which we now know is an important source of scatter in the observed samples.   But this was a promising step forward.

\begin{figure}
\begin{center}
  \includegraphics[width=0.85\textwidth]{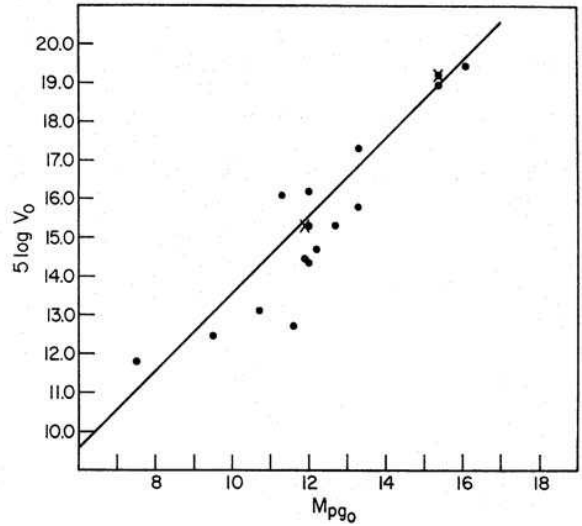}
  \caption{Hubble diagram for 19 SN I from Kowal (1968).}
  \end{center}
  \label{}
\end{figure}

 In 1968, there was plenty of room for improvement in the precision of SN I measurements and in extending the redshift range over which they were studied.  As Kowal forecast: ``These supernovae could be exceedingly useful indicators of distance.  It should be possible to obtain average supernova magnitudes to an accuracy of 5\% to 10\% in the distances.''  He also predicted the future use of supernovae to determine cosmic acceleration:  ``It may even be possible to determine the second-order term in the redshift-magnitude relation when light curves become available for very distant supernovae.''  The ``second-order term'' would be the one that indicated cosmic acceleration or deceleration.  Along with the Hubble constant (which would require reliable distances from Cepheid variables), this deceleration term was expected to provide an account of cosmic kinematics, and, in the context of General Relativity, for the dynamics of the Universe, as sketched for astronomers in the classic paper by Sandage (1961).    

On the last page of this paper,  Sandage worked out the observational consequences of the exponential expansion that would be produced by a cosmological constant. He explicitly shows that you cannot decide between an accelerating universe of this type and the steady-state model (they would both have q$_{0}$ = -1).  Yet, in 1968, the measurement of deceleration was presented by Sandage (1968) as a decisive test between the steady-state model, which predicted acceleration, and Friedmann cosmologies where matter would produce deceleration.  It is possible that, if cosmic acceleration had been discovered earlier, it might have been taken as evidence in favor of the steady-state model.  It was the richer physical context of cosmological information, such as the cosmic microwave background, that led to a much different conclusion in 1998.

\subsection{Classifying supernovae}

In 1968, there was ample room for technical improvement in the measurements themselves, a need for a proper account for the effects of dust, and just as important, well into the 1980s the classification scheme for SN I was still incomplete.  Core-collaspe supernovae were mixed in among the thermonuclear explosions that make up most of the Type I supernovae.  As described by Zwicky (1965) and later by Oke and Searle (1974) the definition of a SN I was empirical: it meant that the spectrum resembled the bright supernova SN 1937C as extensively studied by Minkowski (1939).  The bright supernova SN 1972E, observed with a new generation of spectrophotometric instruments by Kirshner et al. (1973a) in the infrared (Kirshner et al. 1973b) and at late times (Kirshner et al. 1975) provided a rich template for redefining the spectra of Type Ia supernovae.  The distinctive feature in Type I supernova spectra is a broad and deep absorption observed at about 6150 Angstroms, attributed by Pskovskii (1968) to absorption by Si II.  However, there were a handful of SN I, usually dubbed ``peculiar'' SN I, whose spectra resembled the other SN I in other respects, but which lacked this distinctive absorption line at maximum light.  We now understand that this is not just a minor detail: the SN Ib (and their more extreme cousins, the SN Ic) are completely different physical events, ascribed to core-collapse in massive stars that have lost their hydrogen envelopes in late stages of stellar evolution (Branch and Doggett, 1985; Uomoto and Kirshner, 1985; Wheeler and Levreault, 1985;  Wheeler and Harkness, 1990; Filippenko, 1997).  The notation SN Ia was introduced to refer to the original class of supernovae, like SN 1937C and SN 1972E, that has no hydrogen or helium lines in the spectrum and the strong Si II feature.

Once the SN Ib were distinguished from the SN Ia, the homogeneity of the SN Ia improved, with the scatter about the Hubble line decreasing to 0.65-0.36 mag, depending on which objects were selected and which photometric bands were used (Tammann and Leibundgut, 1990; Branch and Miller, 1990, 1993; Della Valle and Panagia, 1992). This work rested on the assumption that the SN Ia were identical, so that a single underlying template for the light curve (Leibundgut, 1988) could be used to interpolate between the observations of any individual object to determine its apparent brightness at maximum light in the B-band, and put all the objects on a common scale.

\subsubsection{SN II as cosmological distance indicators}

The idea that supernovae could be used to measure cosmological parameters had more than one component.  Another line of work employed Type II supernovae.  As pointed out by Kirshner and Kwan (1974), the expanding photospheres of these hydrogen-rich supernovae provide the possibility to measure distances without reference to any other astronomically determined distance.  The idea of the Expanding Photosphere Method (EPM) is that the atmosphere was not too far from a blackbody, so the temperature could be determined from the observed energy distribution.  If you measure the flux and temperature, that determines the angular size of the photosphere.  Since you can measure the temperature and flux many times during the first weeks after the explosion, an observer can establish the angular expansion rate of a supernova.  At the same time, absorption lines formed in the expanding atmosphere, from hydrogen and from weaker lines that more closely trace the expansion of the photosphere, give the expansion velocity.  If you know the angular rate of expansion from the temperature and flux and the linear rate from the shape of the absorption lines, you can solve for the distance to a Type II supernova.  The combination of the supernova's redshift and distance allows for a measurement of the Hubble constant that does not depend on any other astronomically-determined distance.   The departure of the energy distribution for a supernova atmosphere from a blackbody could be computed, as done by Schmidt, Eastman, and Kirshner (1992), and this held out the prospect of making more precise distance measurements to SN II than had been achieved for SN Ia.  

Wagoner (1977) noted that this approach could be extended to high redshift to measure the effects of cosmic deceleration, and also pointed out that the EPM provided an internal test of its own validity: if the distance determined remained the same, while the temperature and the velocity of the atmosphere changed, this was a powerful sign that the measurement was consistent.  This was an important point, since the prospects for using galaxies as the principal tracer of cosmic expansion were dimming, due to evidence that the luminosity of a galaxy could easily change over time due to stellar evolution and galaxy mergers.  Even sign of this change was not known for certain.  Galaxies might grow brighter over time due to mergers, and they might grow dimmer due to stellar evolution.  In either case, unless the effect was carefully calibrated, it could easily swamp the small changes in apparent magnitude with redshift that hold the information on the history of cosmic expansion.   Supernovae, though fainter than galaxies, were discrete events that would not have the same set of changes over cosmic time.  The use of SN II for cosmology has recently been revived and it promises to provide an independent path to measuring expansion and  perhaps even acceleration (Poznanski et al., 2008).

For SN II, the expanding photospheres provide a route to distances that can accommodate a range of intrinsic luminosities and still provide accurate distances, because the atmospheres have hydrogen and behave like those of other stars.   For SN Ia, the atmospheres are more difficult to analyze, but the hope was simpler: that the physics underlying the explosion of a SN Ia would determine its luminosity. The idea that SN Ia were identical explosions has a theoretical underpinning.  In the earliest pictures, the SN Ia were imagined to come from the ignition of a carbon-oxygen white dwarf at the Chandrasekhar mass (Hoyle and Fowler, 1960; Colgate and McKee, 1969).  In models of this type,  a supersonic shock wave travels through the star, burning it thoroughly into iron-peak isotopes, especially Ni$^{56}$.  Such a standard explosion of a uniform mass would lead to a homogenous light curve and uniform luminosity, making  SN Ia into perfect standard candles.  The exponential light curves that suggest an energy input from radioactivity and the late-time spectra of SN Ia, which are made up of blended iron emission lines  were broadly consistent with this picture.  Though the simple theoretical idea that SN Ia are white dwarfs that ignite near the Chandrasekhar mass has been repeated many times as evidence that SN Ia must be perfect standard candles, nature disagrees. Observations show that there is a factor of three range in luminosity from the most luminous SN Ia (resembling SN 1991T) to the least luminous (resembling SN 1991-bg).  Despite the facts, many popular (and professional!) accounts of SN Ia assert that SN Ia are standard candles because they explode when they reach the Chandrasekhar limit.  This is wishful thinking.

\subsubsection{Searching for SN Ia for cosmology }

Nevertheless, the hope that SN Ia might prove to be good standard candles began to replace the idea that brightest cluster galaxies were the standard candles best suited to measuring the deceleration of the universe.  As a coda to his pioneering automated supernova search, Stirling Colgate imagined the way in which a similar search with the Hubble Space Telescope might find distant supernovae (Colgate, 1979).  A more sober analysis of the problem by Gustav Tammann estimated the sample size that would be needed to make a significant detection of deceleration using HST (Tammann, 1979).  The result was encouraging: depending on the dispersion of the SN Ia, he found between 6 and 25 SN Ia at z$ \sim $0.5 would be needed to give a 3$\sigma$ signal of cosmic deceleration.   Tammann got the quantities right-- it was only the sign of the effect that was wrong.

Unwilling to wait for the advent of the Hubble Space Telescope, a pioneering group from Denmark began a program of supernova observations using the Danish 1.5 meter telescope at ESO (Hansen, Jorgensen, and Norgaard-Nielsen, 1987; Hansen et al., 1989).  Their goal was to find distant supernovae, measure their apparent magnitudes and redshifts, and, on the assumption the SN Ia were standard candles, fit for q$_{0}$ from the Hubble diagram.  This method is described with precision in the chapter in this book by Pilar Ruiz-Lapuente.  The difference between q$_{0}$ = 0.1 and q$_{0}$ of 0.5 is only 0.13 mag at redshift of 0.3.  At the time they began their work, there was hope that the intrinsic scatter for SN Ia might be as small as 0.3 mag.   To beat the errors down by root-N statistics to make a 3 sigma distinction would take dozens of well-observed supernovae at z $\sim$ 0.3.  

The Danish group used the search rhythm developed over the decades by Zwicky and his collaborators for finding supernovae.  Since the time for a Type Ia supernova to rise to maximum and fall back by a factor of 2 is roughly one month, monthly observations in the dark of the moon are the best way to maximize discoveries.   Observations made toward the beginning of each dark run were most useful, since that allowed time to follow up each discovery with spectroscopy and photometry.  This is the pattern Zwicky established with the Palomar 18-inch Schmidt and which was used for many years by Sargent and Kowal with the 48-inch Schmidt at Palomar (Kowal, Sargent, and Zwicky, 1970).  It is the pattern used by the Danish group, and all the subsequent supernova search teams until the introduction of dedicated searches like that of Kare et al. (1988) and the rolling search led by John Tonry (Barris et al., 2004) that became the model for the recent ESSENCE and SNLS searches.

But there was something new in the Danish search.  Photographic plates, which are large but non-linear in their response to light, were replaced by a Charge Coupled Device (CCD).  The advantages were that the CCD was much more sensitive to light (by a factor of $\sim$ 100!) and that the digital images were both linear and immediately available for manipulation in a computer.  Fresh data taken at the telescope could be processed in real time to search for new stars, presumably supernovae, in the images of galaxy clusters.  The new image needed to be registered to a reference image taken earlier, the two images appropriately scaled to take account of variations in sky brightness, the better of the two images blurred to match the seeing of the inferior image, and then subtracted.  The Danish team implemented these algorithms and demonstrated their success with SN 1998U, a SN Ia in a galaxy at redshift 0.31 (Norgaard-Nielsen et al., 1989).   Although this group developed the methods for finding distant supernovae in digital data,  the rate at which they were able to find supernovae was disappointingly low.  Instead of making steady progress toward a cosmologically-significant sample at a rate of, say, one object per month, they only found one supernova per year.  At this rate, it would take 10 years to beat down the measuring uncertainty and to begin to learn about the contents of the universe.  And that was in the optimistic case where the intrinsic scatter of SN Ia was assumed to be small. Instead, the observational evidence was pointing in the opposite direction, of larger dispersion among the SN Ia.  Another early effort, carried out by the Lawrence Berkeley Laboratory at the 4m Anglo-Australian Telescope had even less luck.  Despite building a special-purpose prime focus CCD camera to find supernovae, they reported none (Couch et al., 1991).   

\subsection{SN Ia as standard candles-- not!}

Starting in 1986, careful observations made with CCD detectors showed ever more clearly that the luminosity and the light curve shapes for SN Ia were not uniform (Phillips et al., 1987).  In 1991, two supernovae at opposite extremes of the luminosity scale showed for certain that this variety was real, and needed to be dealt with in order to make SN Ia into effective distance measuring tools.  SN 1991bg (Leibundgut et al., 1993; Filippenko et al., 1992) was extremely faint and SN 1991T (Phillips et al., 1992) was extremely bright.  Despite hope for a different result, and a theoretical argument why their luminosities should lie in a narrow range,  Type Ia supernovae simply are not standard candles: they are known to vary over a factor of three in their intrinsic luminosity.  The size of the sample needed to make a cosmological measurement scales as the square of the scatter, so, in 1991, the truly productive thing to harness supernovae for cosmology was not to find more distant supernovae, but to learn better how to reduce the uncertainty in the distance for each object.

Using a set of well-sampled SN Ia light curves with precise optical photometry and accurate relative distances, Phillips (1993) demonstrated  a correlation between the shape of a SN Ia light curve and the supernova's luminosity.  Supernovae with the steepest declines are the least luminous.  More interestingly, even among the supernovae that do not lie at the extremes of the distribution marked by SN 1991T and SN 1991bg, the relation between luminosity and light curve shape provides an effective way to decrease the scatter in the Hubble diagram for SN Ia.  Phillips used this correlation to decrease the observed scatter about the Hubble line to about 0.3 mag.

This made the path forward a little clearer.  What was needed was a well-run supernova search for relatively nearby supernovae that could guarantee accurate follow-up observations.  Mark Phillips, Mario Hamuy, Nick Suntzeff and their colleagues at  Cerro Tololo Inter-American Observatory and at the University of Chile's  Cerro Cal\'an observatory worked together to conduct such a search, the Cal\'an-Tololo Supernova Search (Hamuy et al., 1993).  The technology was a hybrid of the past and the future-- photographic plates were used on the venerable Curtis Schmidt telescope (named in honor of Heber D. Curtis, of the debate cited earlier) at Cerro Tololo to search a wide field (25 square degrees) in each exposure.  Despite the drawbacks of photographic plates as detectors, this large field of view made this the most effective search for nearby supernovae.  The plates were developed on the mountain, shipped by bus to Santiago, and then painstakingly scanned by eye with a blink comparator to find the variable objects.  The modern part was the follow-up.  Since the search area was large enough to guarantee that there would be objects found each month, CTIO scheduled time in advance on the appropriate telescopes for thorough photometric and spectroscopic  follow-up with CCD detectors.  The steady weather at Cerro Tololo and the dedicated work at Cerro Cal\'an led to a stream of supernova discoveries and a rich collection of excellent supernova light curves. For example, in 1996, the Cal\'an-Tololo group published light curves of 29 supernovae obtained on 302 nights in 4 colors (Hamuy et al., 1996a). This is what was needed to develop reliable ways to use the supernova light curves to determine the intrinsic luminosity of SN Ia, and to measure the luminosity distance to each object (Hamuy et al., 1996b).  The Cal\'an-Tololo Search was restricted to redshifts below  0.1, so it did not, by itself, contain information on the cosmology.  However, it provided the data needed to understand how to measure distances with supernovae, and, when used in combination with high-z supernovae, it had the potential to help determine the cosmology.

\subsection{Dust or cosmology?}

However, the accuracy of the distance measurements was compromised by the uncertain amount of dust absorption in the each supernova host galaxy.  Two parallel approaches were developed.  One, led by Mark Phillips and his colleagues, used the observational coincidence, first noted by Paulina Lira, that the evolution in the color B-V had a very small dispersion at ages from 30 to 90 days after maximum (Phillips et al., 1999). By measuring the observed color at those times, the absorption could be inferred and the true distance measured.  The other, based on the same data set, and then later extended through observations at the Whipple Observatory of the Center for Astrophysics, used an empirical method to find that intrinsically faint supernovae are also intrinsically redder.  Since the light curve shape, which was the strongest clue to supernova luminosity, was not greatly affected by absorption, it was possible to determine both the distance and the absorption by dust to each supernova.  A formal treatment of the extinction using Bayes' theorem was used to determine the best values and their uncertainty (Riess, Press, and Kirshner, 1996a).  This MLCS (Multi-color Light Curve Shape) approach was also used to examine whether the dust in other galaxies was the same as dust in the Milky Way (Riess, Press, and Kirshner, 1996b).  While the early indications were that the dust in other galaxies had optical properties that were consistent with those found in the Galaxy, as the samples of supernovae have grown larger and the precision of the measurements has improved, this simple picture is no longer tenable.  These early workers recognized that measuring the extinction to individual supernovae was an essential step in deriving reliable information on the cosmology.  After all, the dimming due to an accelerating cosmology at redshift 0.5 is only of order  0.2 magnitudes.  If instead this dimming were produced by dust like the dust of the Milky Way, the additional reddening would be only 0.07 mag in the B-V color, so good photometry in multiple bands was essential to make reliable inferences on the presence or absence of cosmic acceleration.

\subsection{Early results}

The earliest observations of the Supernova Cosmology Project (SCP) did not take account of these requirements.  Their observations of SN 1992bi at z= 0.458 were made in only one filter, making it impossible, even in principle, to determine the reddening (Perlmutter et al., 1995).  No spectrum for this object was obtained, but it was completely consistent with being a SN Ia.  This was a striking demonstration that the search techniques used by the SCP, which resembled those of the Danish team, could reliably detect transient events in galaxies at the redshifts needed to make a cosmologically interesting measurement.   The search was carried out with a 2048 by 2048 pixel CCD camera at the 2.5 m Isaac Newton Telescope, whose increased speed over the Danish system made it plausible that a supernova could be found in each month's observing.  As with the Cal\'an-Tololo search being carried out at low redshift, it was reasonable for the SCP to schedule follow-up observations.  The SCP developed the ``stretch'' method for accounting for the connection between luminosity and light curve shape in the B and V bands.  This works very well, but does not, by itself, account for the effects of dust extinction (Goldhaber et al., 2001).

The High-Z Supernova Team (HZT) was formed in 1995 by cooperation between members of the Cal\'an-Tololo group and supernova workers at the Harvard-Smithsonian Center for Astrophysics and ESO.  The goal was to apply the new methods for determining the intrinsic luminosity and reddening of a supernova, developed from the low-redshift samples, to objects at cosmologically interesting distances.  This required mastering the techniques of digital image subtraction.  The first object found by the High-Z Team was SN 1995K, at a redshift of 0.479, which, at that time was the highest  yet published (Leibundgut et al., 1996).  Observations were obtained in two colors, and the supernova's spectrum showed it was a genuine Type Ia.  Leibundgut et al. used the observations to show that the light curve for SN 1995K was stretched in time by a factor of (1 + z), just as expected in an expanding universe.  

The time-dilation effect had been discussed in 1939 by Olin Wilson (1939), sought in nearby data by Rust (1974), and by Leibundgut (1990).  Publications by Goldhaber and his colleagues of the SCP (Goldhaber et al., 1996, 2001) show this effect in their data, though the degeneracy between the light curve shape as analyzed by the ``stretch'' method and time dilation requires some (quite plausible) constraints on changes in the supernova population with redshift to draw a firm conclusion.  Another approach to the same problem uses the evolution of the spectra of SN Ia to show in an independent way that the clocks governing distant supernovae appear to run slower by the factor (1 + z) (Foley et al., 2005; Blondin et al., 2008).

In the mid-1990s, important technical developments improved the ability to discover distant supernovae.  At the National Optical Astronomy Observatories, new 2K x 2K CCD systems were implemented at the 4-meter telescopes at Kitt Peak and at Cerro Tololo. In 1997, the Big Throughput Camera (Wittman et al., 1998) became available for general use at the 4 meter telescope at Cerro Tololo.  This 16 Megapixel camera set the standard for distant supernova searches and was employed by both SCP and HZT as they developed the samples that led to the discovery of cosmic acceleration.

But the path to cosmic acceleration was not smooth or straight.   In July 1997, based on 7 objects, the SCP published the first cosmological analysis based on supernovae (Perlmutter et al., 1997).  Comparing their data from z $\sim$0.4, most of which was obtained through just one filter, to the nearby sample from Cal\'an-Tololo (Hamuy et al., 1996a) they found a best value for $\Omega_M$ of 0.88, and concluded that their results were ``inconsistent with Lambda-dominated, low-density, flat cosmologies.''  

Some theorists had begun to speculate that $\Lambda$ was the missing ingredient to reconcile the observations of a large value for the Hubble Constant (Freedman, Madore, and Kennicutt, 1997), the ages of globular clusters, and a low value for $\Omega_M$ in a flat cosmology (Ostriker and Steinhardt, 1995; Krauss and Turner, 1995).  If the universe was flat with a total $\Omega$ of 1, and had $\Omega_M$ of 0.3, then subtraction pointed to a value for $\Lambda$ of 0.7 and you could match the ages of the globular clusters even if the Hubble constant was significantly larger than previously thought. But the initial results of the SCP pointed in the opposite direction, and their evidence for deceleration threw the cold water of data on these artfully-constructed arguments.

The situation began to change rapidly late in 1997.  Both teams used the Hubble Space Telescope to observe supernovae that had been found from the ground.  The precision of the HST photometry was very good, with the supernova well resolved from the host galaxy thanks to the unique angular resolution of HST.  Once the difficult task of accurately connecting the HST photometry to the ground-based work was complete, the observations could be combined to provide additional constraints at the beginning of 1998.  For the SCP, there was one additional object from HST, at a record redshift of  0.83. When combined with a subset of the data previously published in July, the analysis gave a qualitatively different answer.  In their January 1998 Nature paper (submitted on October 7, 1997), the SCP now found that ``these new measurements suggest that we may live in a low-mass-density universe.''  There was no observational evidence presented in this paper for cosmic acceleration (Perlmutter et al., 1998).  For the High-Z Team, the HST-based sample was larger, with 3 objects, including one at the unprecedented redshift of 0.97 (Garnavich et al. 1998).  Although the HZT additional sample of ground-based high-redshift observations was meager (just 1995K), using the same MLCS and template-fitting techniques on both the high-z and low-z samples, and augmenting the public low-z sample from Cal\'an-Tololo with data from the CfA improved the precision of the overall result.  Taken at face value, the analysis in this paper, submitted on October 14, 1997 and published on January 14, 1998, showed the tame result that matter alone was insufficient to produce a flat universe, and, more provocatively,  if you insisted that $\Lambda$ was zero, and the universe was flat, then the best fit to the data had $\Omega_M$ less than 0.  This was a very tentative whisper of what, with hindsight,  we can now see was the signal of cosmic acceleration.

\section{An accelerating Universe}
\subsection{First results}

Both teams had larger samples under analysis during the last months of 1997, and it was not long before the first analyses were published.  The High- Z Team, after announcing their results at the Dark Matter meeting in February 1998 (Filippenko and Riess, 1998), submitted a long article entitled ``Observational Evidence from Supernovae for an Accelerating Universe and a Cosmological Constant'' to the Astronomical Journal on March 13, 1998.  This appeared in the September 1998 issue (Riess et al., 1998).  It used a sample of 16 high-z and 34 nearby objects obtained by High-Z Team members, along with the methods developed at the CfA and by the Cal\'an-Tololo group to determine distances, absorptions, and their uncertainties for each of these objects. The data clearly pointed to cosmic acceleration, with luminosity distances in the high-z sample 10-15\% larger than expected in a low-mass density universe without $\Lambda$.  The High-Z Team also published a long methods paper (Schmidt et al., 1998) and an analysis of this data set in terms of the dark energy equation of state (Garnavich et al., 1998). 

The SCP, after showing their data at the January 1998 AAS meeting, cautiously warned that systematic uncertainties, principally the possible role of dust absorption, made it premature to conclude the universe was accelerating.  They prepared a long paper for publication that showed the evidence from 42 high redshift objects and 18 low-redshift objects from the Cal\'an-Tololo work.  This was submitted to the Astrophysical Journal on September 8, 1998 and appeared in June 1999 (Perlmutter et al., 1999). Although the SCP had no method of their own for determining the reddening and absorption to individual supernovae, they showed that the color distributions of their high-z sample and the objects they selected from the Cal\'an-Tololo sample had similar distributions of restframe color, an indication that the  extinction could not be very different in the two samples.  They also applied the method of Riess, Press, and Kirshner (1996a) to determine the absorption in the cases where they had the required data.  The analysis showed, with about the same statistical power as the High-Z Team paper, that the luminosity distances to supernovae clearly favored a picture in which the universe was accelerating.

\subsection{Room for doubt?}

Two important questions about these results soon surfaced.  

One was whether the results of the two groups were independent.  Some of the machinery for analyzing the data sets, for example, the K-corrections to take account of the way supernova redshifts affect the flux in fixed photometric bands, were based on the same slender database of supernova spectra.  Similarly, the low-redshift sample used by the SCP was made up entirely of objects observed by the HZT.  The two teams cooperated on observing a few of the high redshift objects and both teams used the data for those objects.  A small number of co-authors showed up on both the High-Z Team and the Supernova Cosmology Project publications.  But the analysis was done independently, most of the high-redshift samples were disjoint, and the astronomical community generally took the agreement of  two competing teams to imply that this result was real.  But it was the integrity of the results, not the friction of the personalities, that made this work credible.

Another question about the initial results was whether the measured effect-- a small, but significant dimming of the distant supernovae relative to nearby ones, was due to cosmology, to some form of dust, or to evolution in the properties of SN Ia with redshift (Aguirre, 1999a,b; Aguirre and Haiman, 2000; Drell, Loredo, and Wasserman, 2000).   Aguirre explored the notion that there might be ``grey dust'' that would cause dimming without reddening.  Theoretical difficulties included the limit imposed by using all the available solids, distributing them uniformly, and staying under the limit imposed on the thermal emission from these particles by observations in the far infrared.  A direct approach to the possible contribution of dust came from measurements of supernovae over a wider wavelength range-- the dust could not be perfectly grey, and a wider range of observations, made with infrared detectors, would reveal its properties more clearly.  The earliest application of this was by the HZT (Riess et al., 2000), who observed  a supernova at z = 0.46 in the rest-frame I band, with the goal of constraining the properties of Galactic dust or of the hypothetical gray dust.  They concluded that the observed dimming of the high-z sample was unlikely to be the result of  either type of dust.  Much later, this approach was employed by the SCP (Nobili et al., 2005).  Dust obscuration, and the relation of absorption to reddening, remains the most difficult problem in using supernova luminosity distances for high-precision cosmology, but the evidence is strong that dust is not responsible for the $\sim $0.25 mag dimming observed at z $\sim $0.5.

A second route to excluding grey dust was to extend observations of SN Ia to higher redshift.  If the dimming were due to uniformly distributed dust, there would be more of it along the line-of-sight to a more distant supernova.  Due to the discovery of a supernova in a repeat observation of the Hubble Deep Field (Gilliland, Nugent, and Phillips, 1999) and unconscious follow-up with the NICMOS program in that field, Adam Riess  and his collaborators were able to construct observations of SN 1997ff at the extraordinary redshift of z $\sim $1.7 (Riess et al., 2001).  In a flat universe with $\Omega_{\Lambda} \sim 2/3$ and $\Omega_{M} \sim 1/3$, there is a change in the sign of the expected effect on supernova apparent brightness.  Since the matter density would have been higher at this early epoch by a factor (1+z)$^{3}$, the universe would have been decelerating at that time, if the acceleration is due to something that acts like the cosmological constant.  The simplest cosmological models predict that a supernova at z $\sim$1.7 will appear brighter than you would otherwise expect.  Dust cannot reverse the sign of its effect, so these measurements of the light curve of a SN Ia at z $\sim$1.7 provided a powerful qualitative test of that idea.  While the data were imperfect, the evidence, even from this single object, was inconsistent with the grey dust that would be needed to mimic the effect of cosmic acceleration at lower redshift.  

Another way of solidifying the early result was to show that the spectra of the nearby supernova of Type Ia, the supernovae at z $\sim $0.5 that gave the strongest signal for acceleration, and spectra of the most distant objects beyond z of 1 give no sign of evolution.  While the absence of  systematic changes in the spectra with epoch isn't proof that the luminosities do not evolve, it is a test which the supernova could have failed.  They do not fail this test.  The early HZT results  by Coil et al. (2000) show that, within the observational uncertainties, the spectra of nearby and the distant supernovae are indistinguishable.    This approach was explored much later by the SCP (Hook et al., 2005) with consistent results.

\subsection{After the beginning}

By the year 2000, the context for analyzing the supernova results, which give a strong constraint on the combination ($\Omega_{\Lambda} - \Omega_{M}$) soon included strong evidence for a flat universe with ($\Omega_{\Lambda} + \Omega_{M} = 1$) from the power spectrum of the CMB (de Bernardis et al., 2002) and stronger evidence for the low value of $\Omega_M$ from galaxy clustering surveys (Folkes et al., 1999).  The concordance of these results swiftly altered the conventional wisdom in cosmology to a flat $\Lambda$CDM picture.  But the concordance of these various methods does not mean that they should lean on each other for support like a trio of drunkards.  Instead, practitioners of each approach need to assess its present weaknesses and work to remedy those.  For supernovae, the opportunities included building the high-z sample, which was still only a few handfuls, extending its range to higher redshift, augmenting the low-z sample, identifying the systematic errors in the samples, and developing new, less vulnerable methods for measuring distances to supernovae. 

\subsubsection {Building the High- z sample}

The High-Z Team published additional data in 2003 that augmented the High-Z sample and extended its range to  z = 1.2  (Tonry et al., 2003).  Using the 12K CCD detector at the Canada-France-Hawaii Telescope and the Suprime-Cam at Subaru 8.2m telescope, the HZT then executed a  ``rolling'' search of repeated observations with a suitable sampling interval of 1-3 weeks for 5 months (Barris et al., 2004). This enabled the High-Z Team to double the world's sample of published objects with z $>$ 0.7, to place stronger constrains on the possibility of grey dust, and improve knowledge of the dark energy equation-of-state.  The publication by the SCP of  11 SN Ia with 0.36 $<$ z $<$ 0.86 included high-precision HST observations of  the light curves and full extinction corrections for each object (Knop et al., 2003).

By this point, in 2003, the phenomenon of cosmic acceleration was well established and the interpretation as the effect of a negative pressure component of the universe fit well into the concordance picture that now included results from WMAP (Spergel et al., 2003).  But what was not so clear was the nature of the dark energy.  Increasing the sample near z $\sim$0.5 was the best route to improving the constraints on dark energy.  One way to describe the dark energy is through the equation of state index $w=p/\rho$.  For a cosmological constant, 1 +w = 0.   Back-of-the-envelope calculations showed that samples of a few hundred high-z supernovae would be sufficient to constrain w to a precision of $10\%$.  As before, two teams undertook parallel investigations.  The Supernova Legacy Survey (SNLS), carried out at the Canada-France-Hawaii telescope, included many of the SCP team.  The ESSENCE program (Equation of State: SupErNovae trace Cosmic Expansion) carried out at Cerro Tololo included many of the High-Z Team.  This phase of constraining dark energy is thoroughly described in the chapter in this book by Michael Wood-Vasey,

The SNLS observing program was assigned 474 nights over 5 years at CFHT.  They employed the one-degree imager, Megacam, to search for supernovae and to construct their light curves in a rolling search, with a 4 day cadence, starting in August of 2003.  In 2006, they presented their first cosmological results, based on 71 SN Ia, that gave a value of 1+w = -0.023 with a statistical error of 0.09, consistent with a cosmological constant (Astier et al., 2006).

The ESSENCE program used the MOSAIC II imager at the prime focus of the 4m Blanco telescope.  They observed with this 64 Megapixel camera every other night for half the night during the dark of the moon in the months of October, November, and December for 6 years, starting in 2002.  The survey is described by Miknaitis et al. (2007) and cosmological results from the first 3 years of data were presented in 2006 (Wood-Vasey et al., 2007).  The ESSENCE analysis of 60 SN Ia gave a best value for 1+ w = -0.05, with a statistical error of  0.13, consistent with a cosmological constant and with the SNLS results.  Combining the SNLS and ESSENCE results gave a joint constraint of 1 + w = -0.07 with a statistical error of 0.09.

We can expect further results from these programs, but the easy part is over.  Bigger samples of distant supernovae do not assure improved knowledge of dark energy because systematic errors are now the most important source of uncertainty.  These include photometric errors and uncertainties in the light curve fitting methods,  but also more subtle matters such as the way dust absorption affects the nearby and distant samples.  Collecting large samples is still desirable, especially if the photometric errors are small, but tightening the constraints on the nature of dark energy will also demand improved understanding of supernovae and the dust that dims and reddens them.

\begin{figure}
\begin{center}
\includegraphics[width=0.95\textwidth]{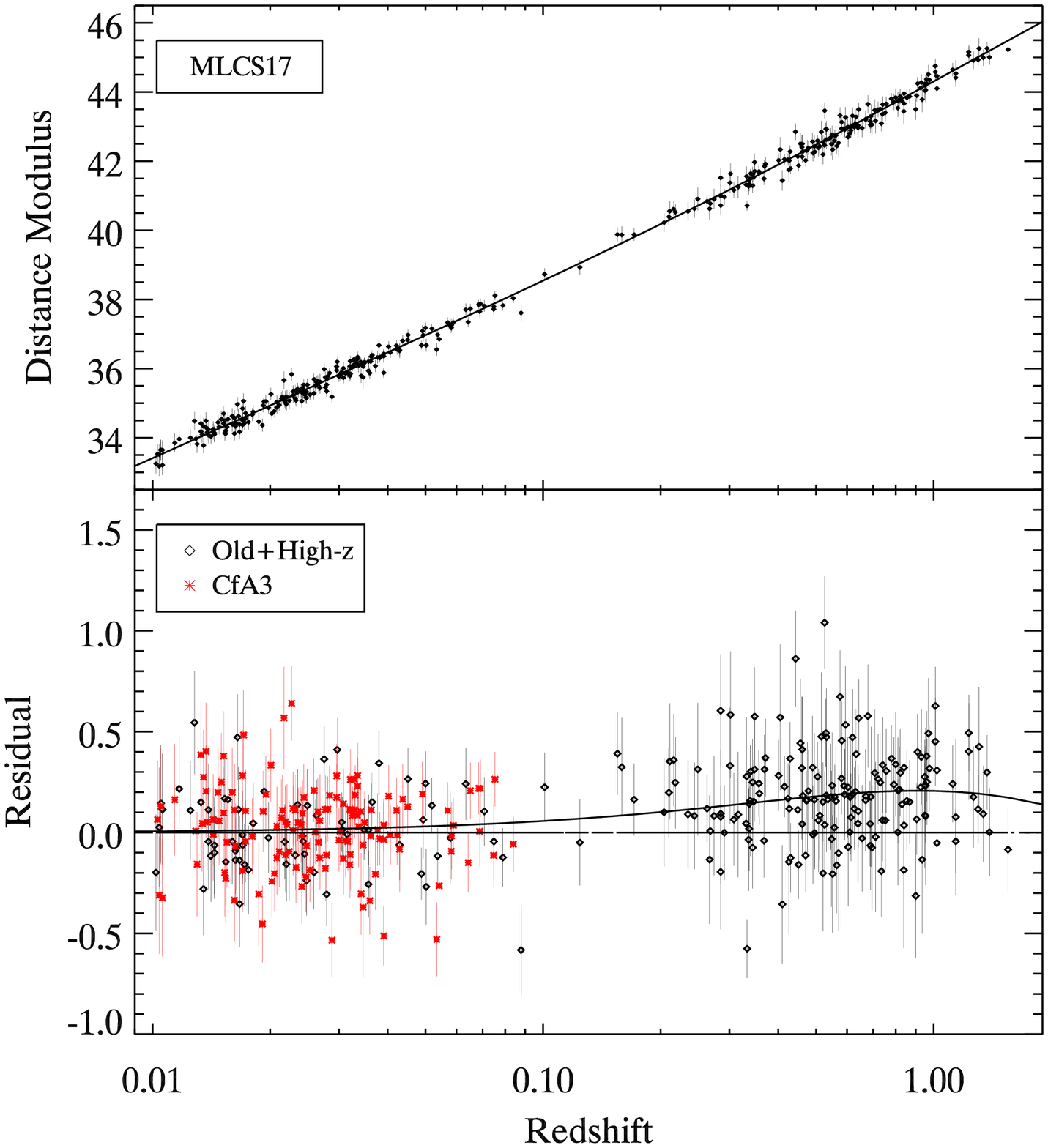}
\includegraphics[width=1.00\textwidth]{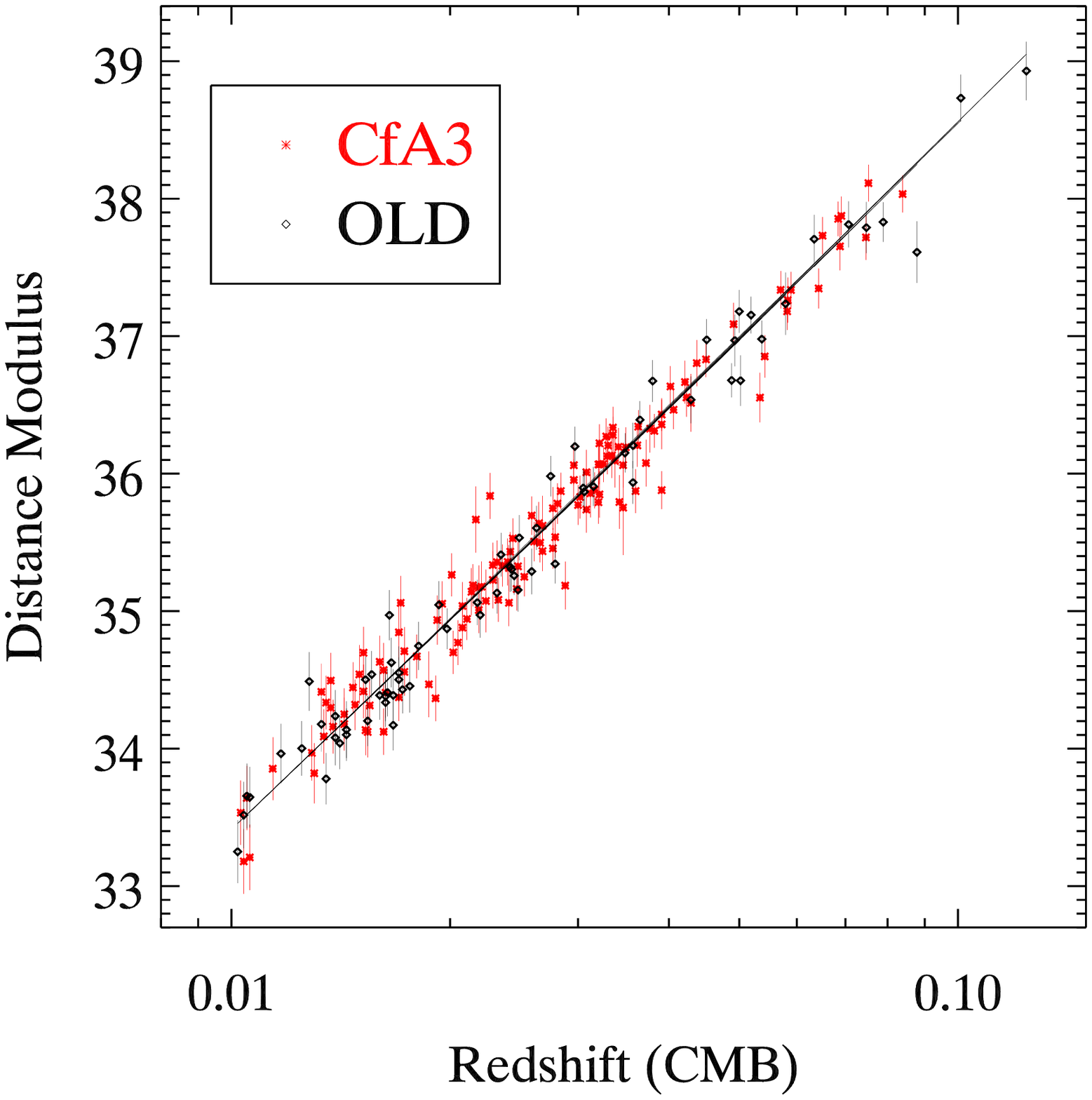}
\caption{{\em Top panels:} Hubble diagram and residuals for MLCS17. The new
CfA3 points are shown as rhombs and the OLD and High--z points as crosses.
{\em Bottom panel:} Hubble diagram of the CfA3 and OLD nearby SN Ia
(from Hicken et al., 2009a).} 
\end{center}
\label{}
\end{figure}

\subsubsection {Extending its range}

While the work of Tonry et al. (2003) and Barris et al. (2004) showed that it was possible, with great effort, to make observations from the ground of supernovae beyond a redshift of 1, the installation of the Advanced Camera for Surveys (ACS) on HST provided a unique opportunity to search for and follow these extremely high-redshift objects (Blakeslee et al., 2003).  By enlisting the cooperation of the GOODS survey, and breaking its deep exposures of extragalactic fields into repeated visits that formed a rolling search, the Higher-Z Team, led by Adam Riess,  developed effective methods for identifying transients, selecting the SN Ia from their colors, obtaining light curves, determining the reddening from  IR observations with NICMOS, and measuring the spectra with the grism disperser that could be inserted into the ACS (Riess et al., 2004a,b, 2007).  This program has provided a sample of 21 objects with z $>$ 1,  and demonstrated directly the change in acceleration, the ``cosmic jerk'' , that is the signature of a mixed dark matter and dark energy universe.  The demise of the ACS brought this program to a halt.  It is possible that the planned servicing mission can restore HST to this rich line of investigation.

\begin{figure}
\begin{center}
\includegraphics[width=0.475\textwidth]{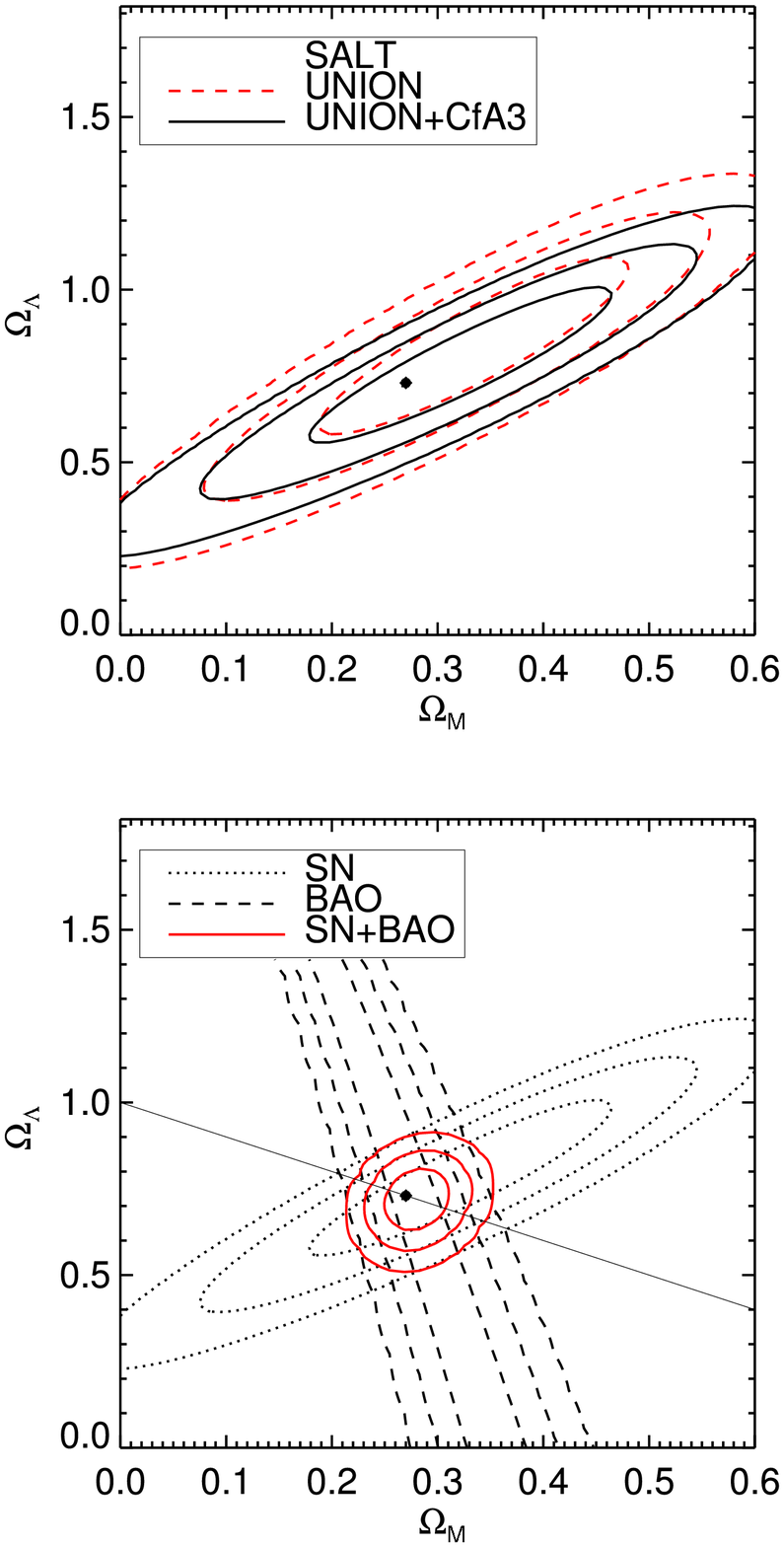}
\includegraphics[width=0.475\textwidth]{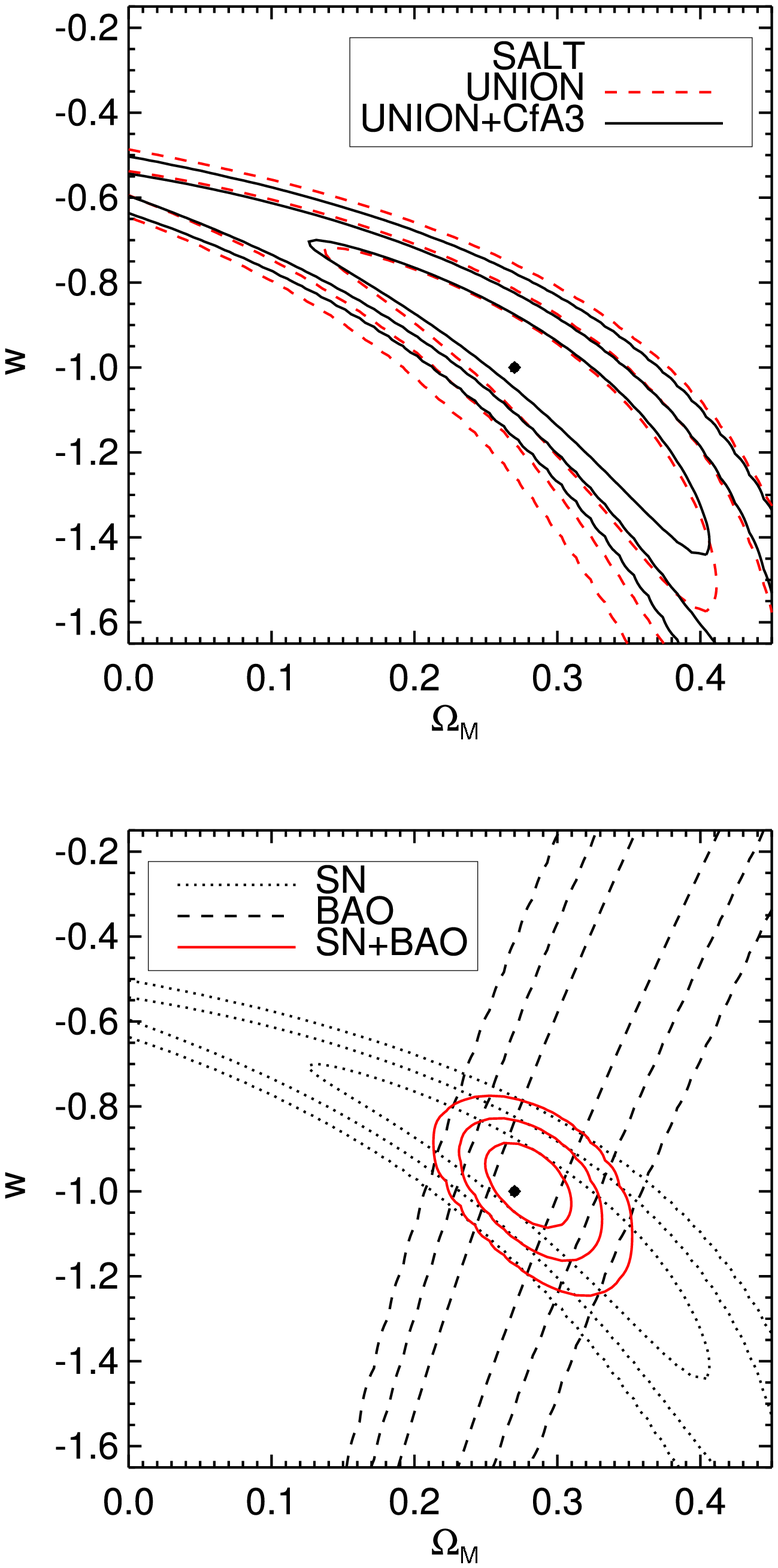}
\caption{{\em Left panels:} Today's best constraints from the Constitution data set on $\Omega_{M}$ and $\Omega_{\Lambda}$. The lower panel shows the 
combination of the SN contours with the BAO prior.
{\em Right panels:} Same for $w$ versus $\Omega_{M}$ in a flat 
Universe
(Hicken et al., 2009b).} 
\end{center}
\label{}
\end{figure}

\subsubsection {Augmenting  the low-z sample}

Both the High-Z Team (which included members of the Cal\'an-Tololo supernova program) and the SCP depended on low redshift observations of supernovae to establish the reality of cosmic acceleration.  The samples at high redshift were assembled, at great effort, and high cost in observing time at the world's largest telescopes because it was clear that these data could shift our view of the universe.  The low-z samples require persistence, careful attention to systematic effects, and promised no shift in world view.  They have been slower to develop.  Two early steps forward were the publication by Riess  of 22 BVRI light curves from his Ph.D. thesis at Harvard (Riess et al., 1999), and the publication of 44 UBVRI light curves from the thesis work of Jha (Jha, Riess, and Kirshner, 2007).  The U-band observations in Jha's work were especially helpful in analyzing the HST observations of the Higher-Z program, since, for the highest redshift objects observed with HST, most of the observations correspond to ultraviolet emission in the supernova's rest frame. Jha also revised and retrained the MLCS distance estimator that Riess had developed, using this larger data set, and dubbed it MLCS2k2. Recently, Kowalski compiled the ``Union'' data sample (Kowalski et al., 2008).  His work assessed the uncertainties in combining data from diverse sources, and, by applying stringent cuts to the data, provided a set of 57 low-redshift and 250 high-redshift supernovae to derive constraints on dark energy properties.  Kowalski noted the imbalance of the low-z and high-z samples and emphasized the opportunity to make a noticeable improvement in the constraints on dark energy by increasing the sample size for the nearby events.  

A third Ph.D. thesis at Harvard, by Malcolm Hicken, has just been completed that finally brings the low-redshift sample out of the statistical limit  created by our slow accumulation of nearby objects and begins to encounter the systematic limit imposed by imperfect distance estimators.  Hicken analyzed the data for 185 SN Ia in 11500 observations made at the Center for Astrophysics over the period from 2001 to 2008.  This large and homogenous data set improves on the Union data set compiled by Kowalski to form the (more perfect) Constitution data set (Hicken et al., 2009a,b).  When Hicken uses the same distance fitter used by Kowalski to derive the expansion history and fits to a constant dark energy, he derives 1 + w = 0.013 with a statistical error of about 0.07 and a systematic error that he estimates at 0.11.  As discussed below, one important contribution to the systematic error that was not considered by Kowalski is the range of results that is produced by employing different  light curve fitters such as SALT, SALT2, and MLCS2k2 which handle the properties of dust in different ways.

This CfA work is a follow-up program that exploits the supernova discovery efforts carried out at the Lick Observatory by Alex Filippenko, Weidong Li, and their many collaborators (Filippenko et al., 2001) as well as a growing pace of supernova discoveries by well-equipped and highly motivated amateur astronomers.   Since the selection of the Constitution supernova sample is not homogeneous, information extracted from this sample concerning supernova parent populations and host galaxy properties needs to be handled with caution, but it suggests that even after light curve fitting, the SN Ia in Scd, Sd, or Irregular galaxy hosts are intrinsically fainter than those in Elliptical or S0 hosts, as reported earlier by Sullivan, based on the SCP sample (Sullivan, 2003).  The idea of constructing a single fitting procedure for supernovae in all galaxy types has proved effective, but it may be missing a useful clue to distinct populations of SN Ia in galaxies that are and are not currently forming stars. There may be a variety of evolutionary paths to becoming a SN Ia that produce distinct populations of SN Ia in star-forming galaxies that are not exactly the same as the SN Ia in galaxies where star formation ceased long ago (Mannucci et al., 2005; Sullivan et al., 2006; Scannapieco and Bildsten, 2005).  Constructing separate samples and deriving distinct light curve fitting methods for these stellar populations may prove useful once the samples are large enough.

A step in this direction comes from the work at La Palma, building up the sample at the sparsely-sampled redshift range near z = 0.2. (Altavilla et al., 2009). A comprehensive approach to sampling has been taken by the Sloan Supernova Survey (Frieman et al., 2008).  By repeatedly scanning a 300 deg$^2$ region along the celestial equator, the survey identified transient objects for spectroscopic follow-up with excellent reliability and has constructed ugriz light curves for over 300 spectroscopically confirmed SN Ia.  With excellent photometric stability, little bias in the supernova selection, and a large sample in the redshift range 0.05 $<$ z $<$ 0.35, this data set will be a powerful tool for testing light curve fitting techniques, provide a low-redshift anchor to the Hubble diagram, and should result in a more certain knowledge of dark energy properties.

In the coming years, comprehensive results from the SCP's	SN Factory (Aldering et al., 2002), the Carnegie Supernova Program (Hamuy et al., 2006), and the analysis of the extensive KAIT archive (Filippenko et al., 2001) should change the balance of the world's sample from one that is just barely sufficient to make statistical errors smaller than systematic errors, to one that provides ample opportunity to explore the ways that sample selection might decrease those systematic errors.

\section {Shifting to the infrared}
	
Coping with the effects of dust absorption was an important contribution of the early work by Phillips et al. (1999) and by Riess, Press, and Kirshner (1996a), the later work by Knop et al. (2003) and Jha, Riess, and Kirshner (2007) and it continues to be the most difficult and interesting systematic problem in supernova cosmology.  The formulations that worked sufficiently well to measure 10\% effects will not be adequate for the high precision measurements that are required for future dark energy studies.   The analysis of the low-redshift supernova data by Conley et al. (2007) showed that either the ratio of reddening to extinction in the supernova hosts was distinctly different from that of the Milky Way (R$_{V}$ = 1.7 instead of the conventional value of 3.1) or there was a ``Hubble Bubble''-- a zone in which the local expansion rate departed from the global value.  As discussed by Hicken (2009a), today's larger sample does not show evidence for the Hubble Bubble, but the value of R$_{V}$ that performs best for MLCS2k2 and for SALT is significantly smaller than 3.1.  It seems plausible that the sampling for earlier work was inhomogeneous, with highly reddened objects present only in the nearby region.  If the correction for reddening in these cases was not carried out accurately, they could contribute to the illusion of a Hubble Bubble.  But the evidence for a small effective value of R$_{V}$ has not gone away.  It seems logical to separate the contribution due to reddening from the contribution that might result from an intrinsic relation between supernova colors and supernova luminosity, as done in MLCS2k2, but the approach of lumping these together, as done by the fitting techniques dubbed SALT and SALT2, also works well empirically (Guy et al., 2005, 2007).  In the ESSENCE analysis, the effects of extinction on the properties of the observed sample were carefully considered, and found to affect the cosmological conclusions.  Getting this problem right will be an important part of preparing for higher precision cosmological measurements with future surveys.

\begin{figure}
\begin{center}
\includegraphics[width=1.00\textwidth]{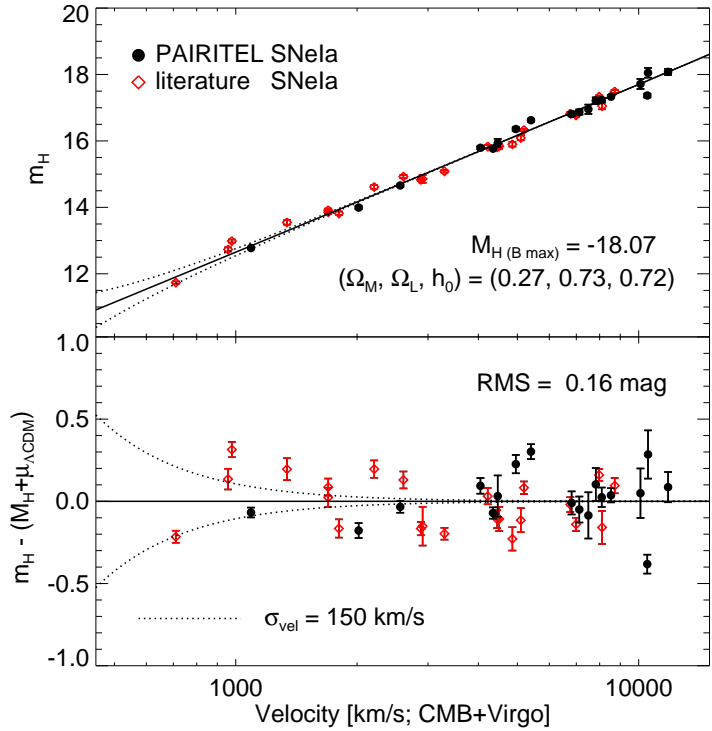}
\caption{$H$--band SN Ia Hubble diagram. It includes 23 new SN Ia observed 
with PAIRITEL (from Wood-Vasey et al., 2008).} 
\end{center}
\label{}
\end{figure}

Fortunately, there is a very promising route to learning more about dust, avoiding its pernicious effects on supernova distances, and deriving reliable and precise measures of dark energy properties.  That route is to measure the properties of the supernovae in the rest frame infrared.  As shown in the pioneering work of Krisciunas, Phillips, and Suntzeff (2004), nearby SN Ia in the Hubble flow behave as  very good standard candles when measured in near infrared bands (NIR), typically J, H, and K$_s$.  This work has recently been extended by Wood-Vasey et al. who used the PAIRITEL system (a refurbished and automated version of the 2MASS telescope) at Mount Hopkins to obtain near infrared light curves that double the world's sample (Wood-Vasey et al., 2008) .  Even with no correction for light curve shape or dust absorption, the NIR light curves for SN Ia exhibit a scatter about the Hubble line that is typically 0.15 mag.  This is comparable to the scatter that is achieved by the output of the elaborate light curve fitters now in use for optical data that correct for the width of the light curve's peak and use the optical colors to infer dust corrections.  This means that the SN Ia actually do behave like standard candles-- but in the NIR!  What's more, the effects of dust absorption generally scale as $1/\lambda$, so the effects of extinction on the infrared measurements should be 4 times smaller than at the B band.  When combined with optical data, the infrared observations can be used to determine the properties of the dust, and to measure even more accurate luminosity distances.  Early steps toward these goals are underway (Friedman et al., 2009).

\section{The next ten years}

Goals for the coming decade are to improve the constraints on the nature of dark energy by improving the web of evidence on the expansion history of the universe and on the growth of structure through gravitation (Albrecht et al., 2006, 2009; Frieman, Turner, and Huterer, 2008; Ruiz-Lapuente, 2007).  Supernovae have an important role to play because they have been demonstrated to produce results.  Precise photometry from homogeneous data, dust absorption determined with near-IR measurements, and constructing useful subsamples in galaxies with differing star formation histories are all areas where we know improvement in the precision of the distance measurements is possible.  More speculative, but plausible, would be the use of supernova spectra in a systematic way to improve the distance estimates.  Implementation of statistically sound ways to use the light curves (and possibly spectra) to determine distances should make the results more reliable and robust.  What is missing is a level of theoretical understanding for the supernova explosions themselves that could help guide the empirical work, and provide confidence that stellar evolution is not subtly undermining the cosmological inferences (Hoeflich, Wheeler, and Thielemann, 1998; Ruiz--Lapuente, 2004).  Large samples from Pan-STARRS, the Dark Energy Survey, and, if we live long enough, from JDEM and LSST will eventually be available.  The chapter in this book by Alex Kim makes a persuasive case for the effectiveness of a thorough space-based study of supernovae.  Our ability to use these heroic efforts effectively depends on  improving our understanding of supernovae as astronomical objects in the context of galaxy formation, stellar evolution, and the physics of explosions. Then we can employ the results with confidence to confirm, or, better yet, to rule out some of the weedy garden of theoretical ideas for the dark energy described in other chapters of this book!   

\section {Acknowledgements}
Supernova research at Harvard is supported by the US National Science Foundation through grant AST06-06772.  I am grateful for the long series of excellent postdocs and students who have contributed so much to this work.  As postdocs, Alan Uomoto,  Bruno Leibundgut, Pilar Ruiz-Lapuente,  Eric Schlegel,  Peter Hoeflich, David Jeffery, Peter Garnavich, Tom Matheson, Stéphane Blondin, and Michael Wood-Vasey.  And as graduate students Ron Eastman, Chris Smith, Brian Schmidt, Jason Pun, Adam Riess, Saurabh Jha, Marayam Modjaz, Malcolm Hicken, Andy Friedman, and Kaisey Mandel.

\end{document}